# Microphase separation as the cause of structural complexity in 2D liquids

Alexander Z. Patashinski,*[a,b] Rafal Orlik,[c] Antoni C. Mitus,[d]
Mark A. Ratner,[a] Bartosz A. Grzybowski,*[a,b]

[a]Department of Chemistry, Northwestern University, 2145 Sheridan Road, Evanston, IL, 60208-3113, USA. E-mail: a-patashinski@northwestern.edu;
[b]Department of Chemical and Biological Engineering, Northwestern University, 2145 Sheridan Road, Evanston, IL, 60208-3113, USA.
E-mail: grzybor@northwestern.edu;
[c]Orlik Software, ul. Lniana 22/12, 50-520 Wrocław, Poland;
[d]Institute of Physics, University of Technology, Wybrzeze Wyspianskiego 27, 50-370 Wroclaw, Poland.

Under certain thermodynamic conditions, a two-dimensional liquid becomes a statistically stable mosaic of small differently-ordered clusters. We apply to this mosaic a special coarsening procedure that accounts for short-time average and topologic features of a particle near environments. We then show that the coarsened mosaic consists of two different components separated at the length-scale of few inter-particle distances. Using bond order parameters and bond lengths as instant local characteristics, we show that these components have internal properties of spatially heterogeneous crystalline or amorphous phases, so the coarsened mosaic can be seen as a microphase-separated state. We discuss general conditions favouring stability of the mosaic state, and suggest some systems for searching for this special state of matter.

Many pure low molecular weight liquids behave as "simple" liquids (fluids) [1-4]. An ideal simple liquid is characterized by a microscopically short internal relaxation time. To a good approximation, liquefied perfect gases (Ar, Cr, etc), many molten metals and salts, and also "theoretical" three-dimensional (*3D*) liquids of Lennard-Jones particles, hard and soft spheres are simple liquids. Scattering experiments and computer simulations show that particles positions in these liquids are loosely correlated at distances around the first peak in the radial distribution function, and all correlations vanish at larger distances. Microscopic theories assuming this simple picture explain and predict properties of simple liquids.

Not included in the list of simple liquids are macromolecular and polymeric liquids, supercooled glassforming liquids near glass transition, colloidal liquids, emulsions and suspensions. These liquids are distinguished by wide spectra of internal relaxation times, non-exponential relaxation kinetics, and other common properties not found in simple fluids, and constitute the class of complex liquids, also referred to as "soft matter" [1-4]. Common features of complex liquids can be understood based on the assumption that particles in these liquids form temporary aggregates so that at any time the liquid is a mosaic [5] of aggregates and non-aggregated clusters; this assumption is supported by observations of tracer diffusion [6,7]. Complex liquids are the subject of intense ongoing research.

The defining feature of the mosaic state is coexistence of structurally different small-size domains. Possibility for a condensed molecular system to exist in more than one macroscopic structural state is the base of the theory of phase transitions [8]. As a rule, the entire volume of a system at equilibrium is occupied by one of the phases (structures), but under special conditions (for example, at phase transition temperature at a constant volume) the system can be in a stable state of phase coexistence in which different phases occupy macroscopic parts of the volume. Here, the notion of stability is important: an unstable mosaic-like state with very small domains of one phase in the matrix of the other can be created, for example, by a rapid change of temperature or volume making the liquid metastable. In the metastable liquid, supercritical nuclei of the new phase appear but these nuclei then grow or coalesce into macroscopic phases [8]. A change of parameters over the spinodal line triggers spinodal decomposition process [9,10] producing small domains of the new phase, but then these domains also grow and coalesce. These and other examples describe unstable mosaics; are there also stable mosaic states?

To test structural stability of a particular complex liquid, one needs to track configurations with molecular-scale resolution – a task currently too challenging for experiment or computer simulation because complex liquids are *3D* molecular systems with particles interacting via sophisticated potentials. Pure *3D* liquids made of point-like particles with simple two-particle interactions are simple liquids up to the crystallization point. In contrast, recent studies of supercooled [11] and equilibrium two-dimensional (*2D)* Lennard-Jones liquids [12,13] found that properties of these liquids substantially differ from their *3D*-analogs. Computer simulations [12], [13], [16-18] confirm the assumption of the theory [14,15] that very close to the melting point *2D* liquids represent a locally crystalline matrix with some concentration of spatially isolated islands of disorder around dislocations. On further increase of the temperature, these islands gradually increase in size and number, and at some temperatures percolate creating a mosaic of crystalline and non-crystalline clusters. At these temperatures the *2D* liquid shows signature features of a complex liquid: super-Arrhenius increase of relaxation times on cooling, a wide spectrum of internal relaxation times, and stretched-exponential relaxation kinetics[12]. *2D* liquids offer an opportunity to study structural heterogeneity in a relatively simple system, with the ultimate hope of obtaining information that could illuminate the general case. As a first step, we study here internal characteristics of small domains comprising the mosaic, and then compare these characteristics to those in uniform liquid and crystalline phases. The surprising result of the study is that in each domain the density, spatial organization, and statistical distributions of local parameters strongly resemble either the spatially-homogeneous crystal or the structureless liquid phase. The *2D* mosaic states are stable at time-scales that are substantially longer than the lifetimes of a single domain.

For this study, the *2D* system of $N = 2500$ Lennard-Jones (*LJ*) particles was simulated under periodic boundary conditions using a rather standard Molecular Dynamics procedure (see, for example, [13]). We tested that increasing the size of the system to $N=16384$ does not result in qualitative changes of behavior. The Lennard-Jones potential acting between every two particles separated by a distance $r$ was chosen in the non-dimensional form

$$\begin{cases} U(r) = F(r) - F(r_{cutoff}), \\ U(r) = 0, \ r > r_{cutoff} \end{cases}$$
$$F(r) = 4\left[\left(\frac{1}{r}\right)^{12} - \left(\frac{1}{r}\right)^{6}\right]$$
, (1)

the cut-off length was $r_{cutoff} = 2.5$. The form (1) of the interaction potential introduces natural units of length, energy, and particles number density $\rho$; the natural unit of temperature $T$ coincides with that of energy (the Boltzmann constant $k_B=1$) [13,15]. Equilibrium states along the super-critical isotherm $T = 0.700$ were simulated in the range of densities $\rho = 0.60$–$0.90$ that includes both the crystalline states at high-density end and non-mosaic liquid states at lowest densities.

A configuration of the system can be seen as a bond network [19] where particles serve as nodes and the lines (bonds) connecting a particle to its six nearest neighbors as edges. These six nearest neighbors of a particle are referred to as its cage particles. In computer simulations, each particle is assigned a unique identifier $a = 1, 2, 3, \ldots, N$ that is conserved during simulation. We denote $d_a$ the six vectors connecting a particle $a$ with its cage particles; the shape of this cage can be described by the lengths $|d_a|$ of the bonds and the angles $\varphi(d_a)$ between the bonds and the *x*-axis of a *2D* Euclidean coordinate frame. The bond orientation order parameters [19, 20] are defined by the formula [15]

$$Q_n(a) = \frac{1}{6}\sum_{d_a} e^{in\phi(d_a)} = |Q_n(a)|\exp(i\psi(a)); \qquad 2$$

the sum in (2) is over all six vectors $d_a$. The moduli $|Q_n|$ of the order parameters characterize the shape of a cage independent from its orientation. For a perfectly hexagonal cage, $|Q_6(a)|=1$ and $|Q_n(a)|=0$ for $n \neq 6$; deviations from this ideal form result in $|Q_6(a)|<1$, $|Q_n(a)|>0$. To better detect these deviations, we use, following [13], [21], the linear combination

$$Q^2 = \sum_{m=0}^{6} |Y_{6m}(\frac{\pi}{2},0)|^2 Q_m^2, \qquad 3$$

where $Y_{6m}(\theta,\varphi)$ are the spherical harmonics. For a perfect hexagon, the parameter $Q$ has its maximum value $Q_{max}=0.74$. A study of the correspondence between cage shapes and values of $Q$ (see [21] and references therein) found that cages with $Q > Q_0 = 0.555$ represent fluctuating hexagons; for brevity, centers of these cages are mentioned below as *SL* (solid-like [21]) particles. Cages with $Q < Q_0$ are distorted to the level when association with regular hexagon becomes ambiguous; centers of these cages are mentioned as *LL* particles. *LL* particles are found in the cores of packing defects (dislocations or vacancies).

It was found in [12] that particles in large *SL* clusters vibrate for many vibration periods without changing their nearest neighbors except for rare and reversible *SL-LL* switches, while in the parts where *LL* and *SL* particles are mixed, *SL-LL* switches are very frequent events. Particles of a perfectly crystalline cluster can be mapped onto sites of a hexagonal lattice. For a crystalline cluster, a one-to-one mapping exists that maps the six nearest neighbors of a particle onto nearest neighbors of this particle's image on the lattice: this mapping conserves the topology of the cluster defined as the nearest-neighbors relations of all cluster's particles.

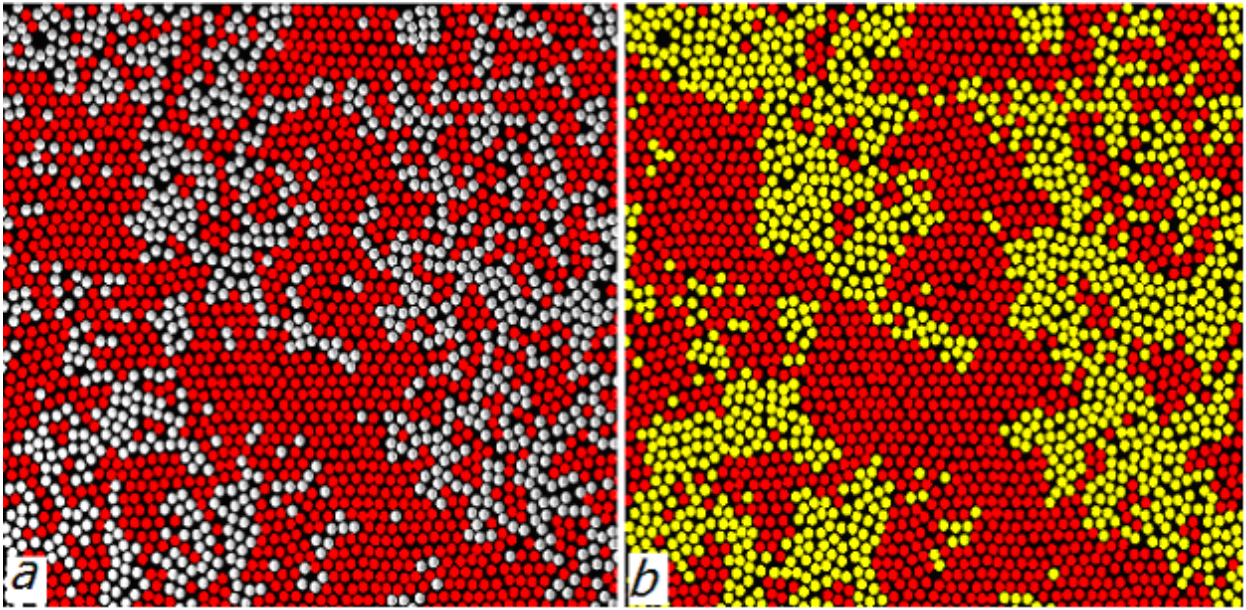

**Figure 1. The mosaic at $T=0.700$, $\rho=0.835$ before (1a) and after (1b) coarsening.**

Tracking changes in these relations shows [13] that reversible *SL-LL* switches represent large and rare displacement events in a perfectly crystalline cluster. Although these events can be expected as part of the thermal motion in a crystal, the apparent violation of the crystalline order in large crystalline clusters is an artifact of the definition [12,13, 19-21] of nearest neighbors because this definition is based on instant bond lengths. A physically correct definition of nearest neighbors has to account for the short-time history and connectivity of the bond network in a volume larger than the cage. The *LL-SL* picture of the system is too local in space and time and needs to be coarsened (short-time-smoothed) to account for the short-time dynamics of the structure in regions larger than the cage; the coarsening procedure has to re-define nearest neighbors of a particle based on topologic rather than metric criteria [22]. In particular, coarsening has to re-define crystalline clusters as topologic (as verified by mapping onto hexagonal lattice) long-living crystallites. We will refer to particles of these crystallites as *CL* particles. One expects that almost all *SL* particles in a large *SL* particles aggregate will be identified as *CL* particles, but also some *LL* particles may be recognized as *SL* particles. The part of the system not belonging to topological crystallites represent non-crystalline, amorphous material; particles in this part will be referred to as *AL* particles. One expects that coarsening will recognize as *AL* particles most of the *LL* particles and also short-living *SL* particles outside the coarsened crystalline clusters.

Manual mapping of configurations onto hexagonal lattice is a time-consuming task, so one would like to teach the computer to identify *CL* and *AL* particles using the abovementioned criteria. While manually performing this task for selected configurations, we observed – and then verified – that most *CL* particles have no more than two *LL* nearest neighbors while most *AL* particles have more than two *LL* nearest neighbors. The identification of *CL* and *AL* particles based on this rule is a coarsening procedure that involves distances about twice the cage size, by using characteristics of nearby cages. This procedure corrects most of the issues mentioned: it does not significantly change large aggregates of *SL* particles except for adding a few occasional internal *LL* particles reassigned as *CL* particles, but it clearly redefines the amorphous clusters. In principle, the procedure can be refined to improve recognition of *CL* and *AL* particles, but already the first iteration eliminates most issues and appears sufficient for statistical studies. Fig.1b shows the crystalline (*CL*) and amorphous (*AL*) aggregates that appear after coarsening the *SL-LL* configuration presented in Fig.1a.

Now that the liquid is represented as a mosaic of crystalline and amorphous regions, one can sample a representative ensemble of bond lengths and bond orientation order parameters in these regions. For each (*CL* and *AL* ) part of the mosaic, the average bond length monotonously increases with decreasing $\rho$ (see Fig. 2), but this increase is slower than that of the system-average bond length (also shown in Fig. 2) because the system-average bond length changes

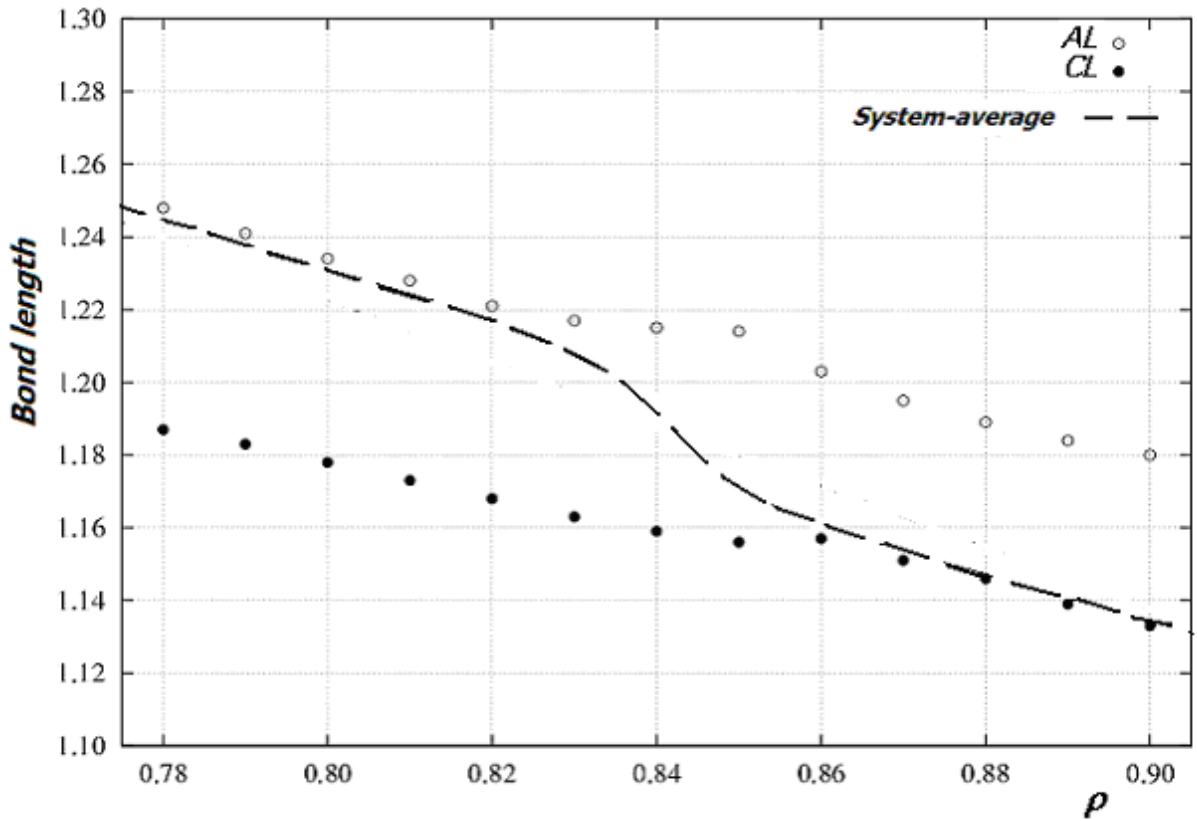

Figure 2. The average bond lengths in *AL* and *SL* clusters.

mainly due to the change in the fractions of *CL* and *AL* components (see Fig. 3). The probability distribution functions (*pdf*) for *CL-CL* and for *AL-AL* bonds length are presented in Fig. 4. Not included in these distributions are *CL-AL* bonds belonging to the crystallites-amorphous interface; the number of these bonds in mosaic states is not small. The functions in Fig. 4 for very short bonds are universally determined by strong repulsion of particles; for longer bonds, data for *CL* clusters are qualitatively different from those for *AL* clusters. In crystalline clusters at $\rho = 0.82$–$0.85$, probability distribution functions shear common qualitative features that are also common with those in the crystal at higher densities $\rho > 0.90$. Surprisingly, probability distribution functions for all amorphous clusters also show common features, in particular a long tail at large distances; these features qualitatively differ from those in *CL* clusters but are

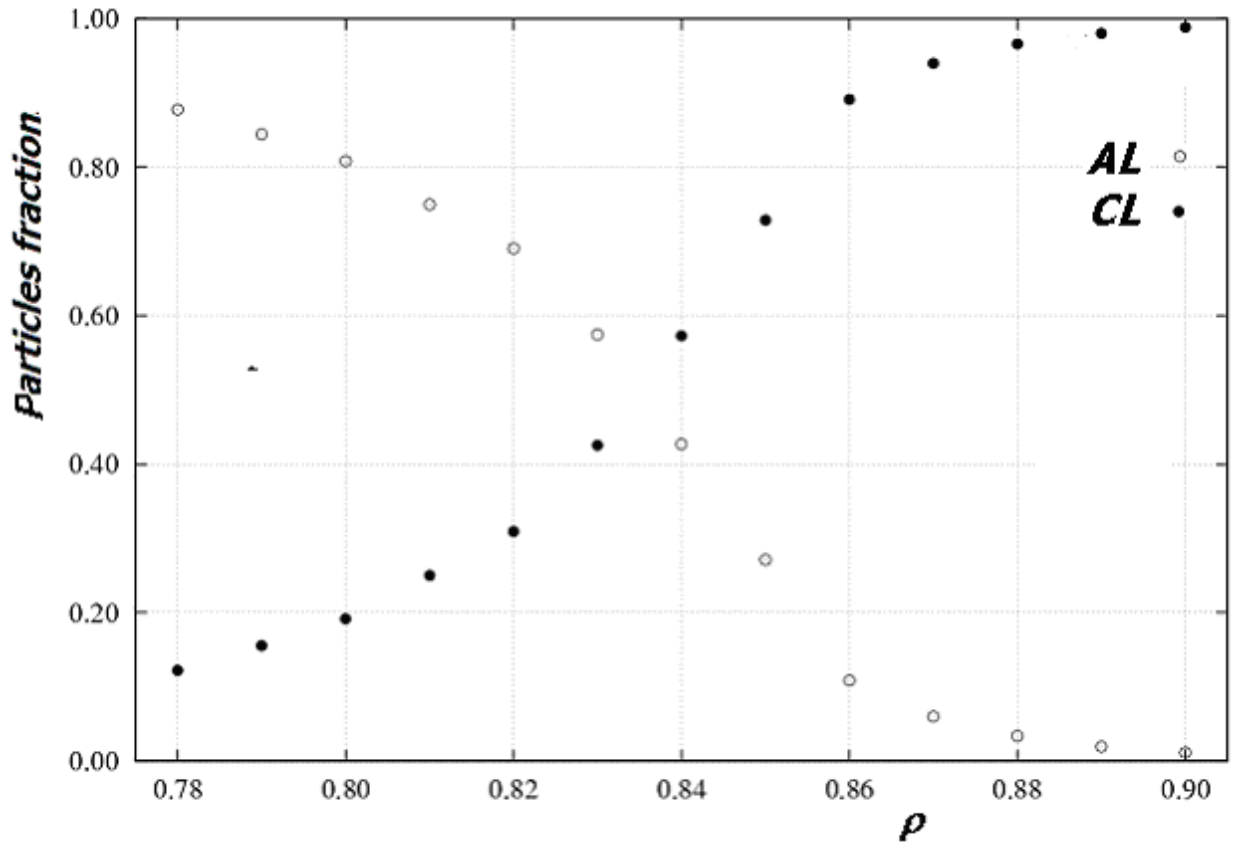

Figure 3. The fractions of *AL* and *CL* particles.

common with those in the structureless liquid at $\rho = 0.60$. One suggests that these common features reflect the nature of the amorphous state: for most particles, one of the six bonds is so long that the association of the corresponding nearest neighbor with the cage is physically ambiguous; a particle is surrounded by only five "real" nearest neighbors. Attempt to map *AL* clusters onto hexagonal lattice shows that a significant concentration of vacancies is necessary to avoid contradictions. The probability distribution function for the bond orientation order parameter $Q$ (Fig. 5) shows quantitative and also qualitative differences between *CL* and *AL* clusters. This function has a high maximum at large values of $Q$ for the crystallites and the

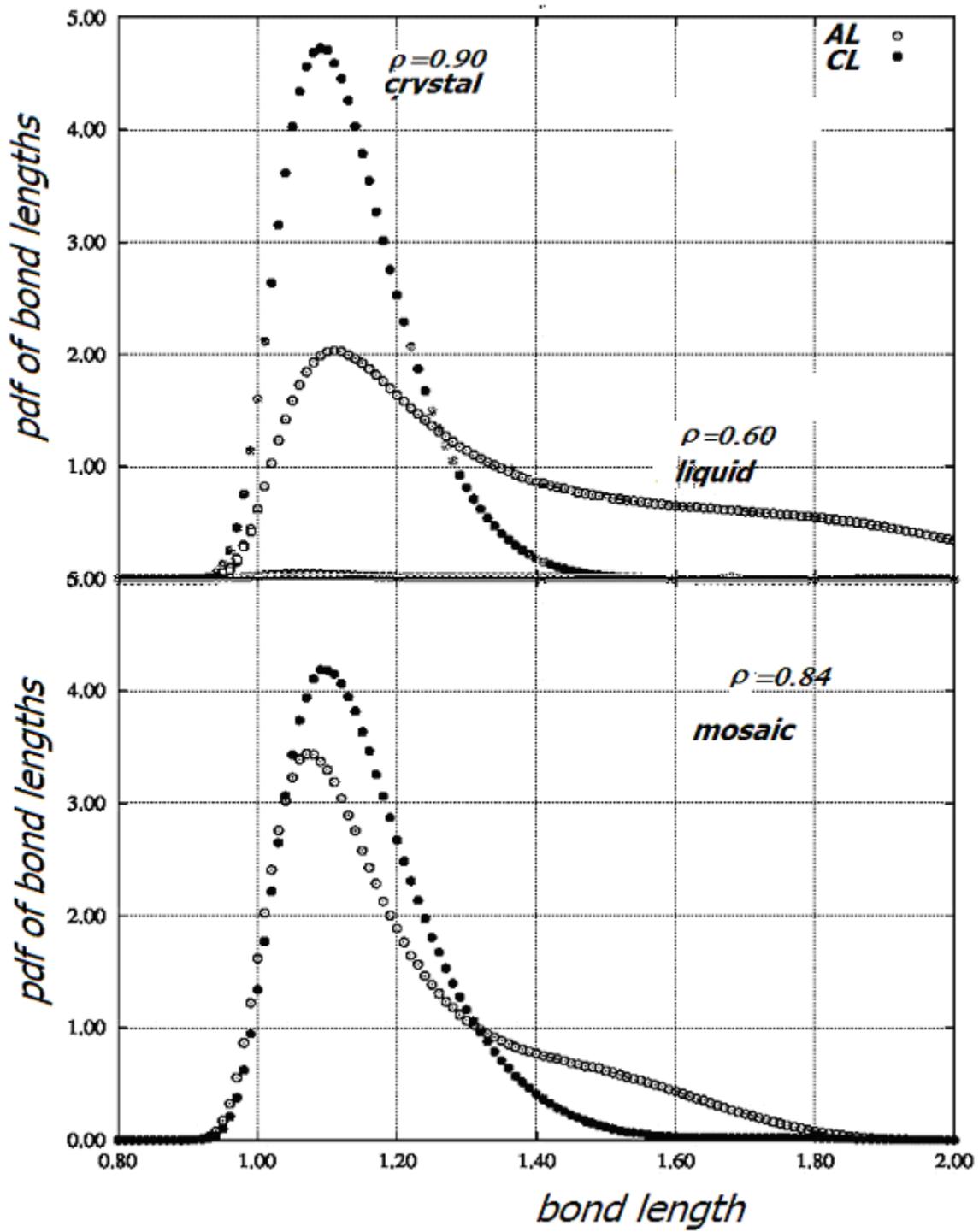

**Figure 4. Bond length probability distributions for uniform phases and for the coarsened**

**mosaic.**

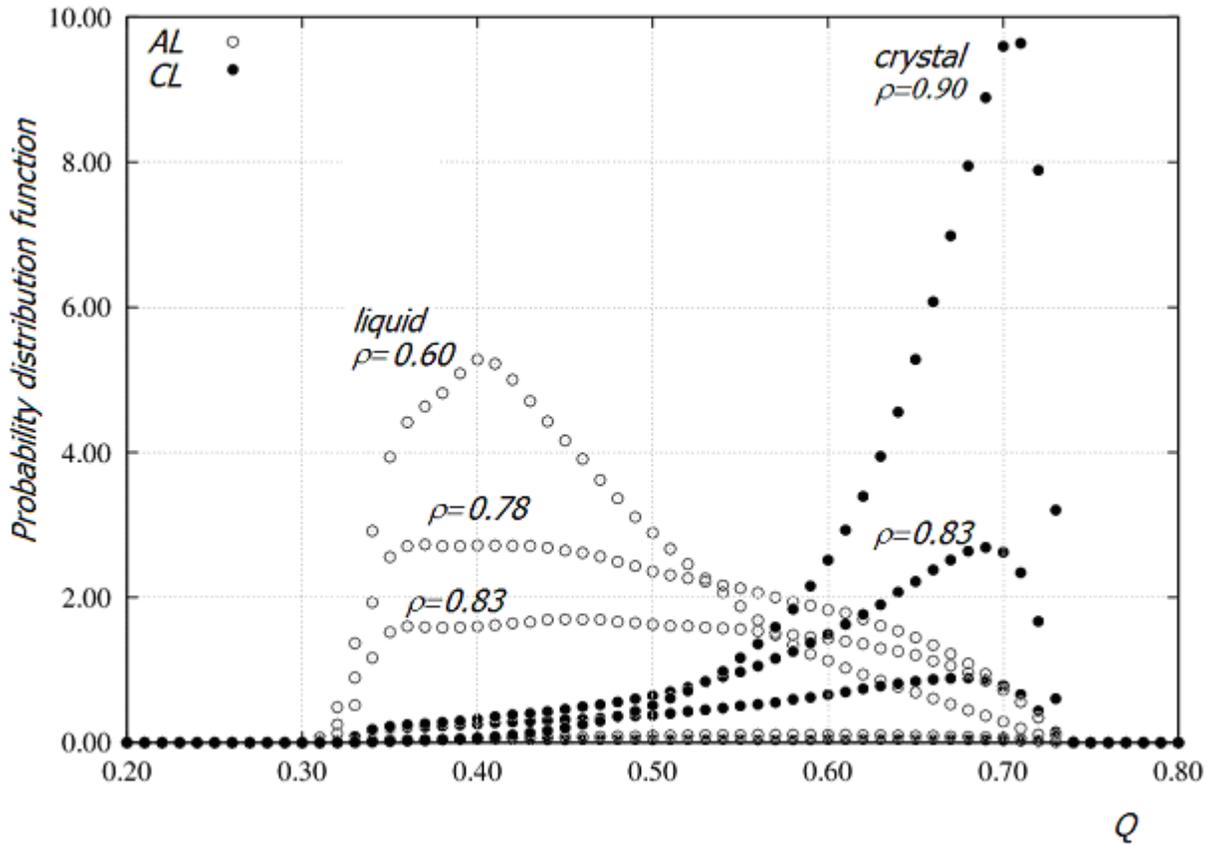

**Figure 5. The probability distribution function the bond orientation order parameter Q for uniform phases and for the coarsened mosaic.**

crystal state, but a much wider profile for the amorphous part. In the amorphous clusters, most of the (*AL*) particles have cages with $Q \sim 0.4$, values that are typical for packing defects with particles surrounded by only five nearest neighbors.

We assume that the remarkable differences in properties between crystallites and amorphous clusters amount to differences between coexisting phases, and the mosaic state in the *2D* liquid can be seen as a stable micro-phase-separated state, a condensed "emulsion" of small (tens or hundreds of particles) clusters of crystalline (solid) phase in a matrix of amorphous (liquid) phase. Unlike emulsions made of stable emulsion particles, the crystallites in the mosaic are self-assembled aggregates having finite lifetimes, and the observed dynamics of self-assembly and disassembly proves that each molecule spends finite times as part of a crystallite or an amorphous cluster [13]. The coarsened picture of the system studied here considers short-time smoothing of fluctuating cages. Large crystallites are not significantly affected by the coarsening procedure; in particular, the longer-time changes in the coarsened mosaic are driven by the micro-melting/micro crystallization processes at the borders of those crystallites described in [12] and [13]. One notes that in the Statistical Mechanics of classical systems, the equilibrium statistics of configurations is independent from their dynamics [8]; once the mosaic is assumed to be the feature of configurations that define the equilibrium ensemble for the liquid, crystallites can be treated as emulsion particles.

Theoretical ideas of particles in liquids forming clouds [23], essential [24], inherent [25], local [26] structures have been around for a rather long time. A mosaic state assumes that there is more than one such structure in the same liquid. As already mentioned, under certain conditions, equilibrium coexistence of different structural states in a molecular system (at polymorphous transitions in crystals, crystal-liquid and liquid-gas coexistence [8], different liquid states in liquid-liquid phase transitions [26-30]) is rather a rule than exception. In macroscopic two-phase states, same-structure domains are merged into macroscopic volumes. A mosaic state yields the necessary condition of each small cluster being in one of the structural states, but same-structure clusters are not forming macroscopic regions so the system is in a micro-phase separated state that, in principle, can be unstable, metastable, or, as the above study of the *2D* Lennard-Jones liquid suggest, thermodynamically stable state. Below, we discuss the stability of micro phase-separated states in the *2D* liquid, and also in other systems.

According to the theory of dislocation-mediated melting of *2D* crystals [14,15], dislocations in the infinite-size crystal form aggregates (pairs, 3-dislocation aggregates, etc) characterized by zero total Burgers vector. The contribution of an aggregate of a size *R* to the free energy of the crystal is $\Delta F(R) \sim \mu(R) \ln R - T \ln R = \{\mu(R) - T\} \ln R$. In this sum, the first term $\Delta F_U \sim \mu(R) \ln R$ describes the elastic energy of the strain caused by the dislocations, and the part $\Delta F_S \sim -T \ln R$ accounts for the entropy related to dislocations positions. Due to the screening of the dislocations by aggregates of smaller sizes, the effective elastic modulus $\mu(R)$ decreases with R increasing. At temperatures below the unbinding temperature $T_{un}$ the energy $\Delta F(R) > 0$ for all *R*, so any defects increase the free energy and this increase is largest for large *R*, so the concentration of aggregates is small and decreases with increasing R. Above the unbinding temperature, the energy $\Delta F(R)$ is positive only for $R < r_{scr}$. To decrease the free energy, aggregates of sizes $R > r_{scr}$ multiply and create some concentration of free dislocations, the topological long-range crystalline order disappears (although the orientation order may decay only algebraically if the system enters the hexatic phase [14,15,17].

Visualization of configurations just above the expected melting temperature confirms the picture of a crystalline matrix with a small number of isolated dislocations; in this state, the finite-size system has significant orientation order [13]. The picture changes (see Supplemental material, available online) when the temperature is gradually increased or density decreased: small islands of disordered material appear surrounding the dislocations and then grow in size and number until an almost percolated or percolated network of disordered (amorphous, *AL*) material appears; this is the mosaic state. The mosaic state cannot be described by asymptotic formulas of the unbinding theory. However, these formulas are applicable to a hypothetic crystallite of a size $R \gg r_{scr}$. In qualitative agreement with the unbinding theory for infinite systems, screening effects are expected to result in proliferation of dislocation aggregates of sizes that are larger than the screening length $r_{scr}$ but smaller than the size of the crystallite, turning the hypothetic large aggregate into mosaic of smaller crystallites. Then, the sizes of large crystallites can not substantially exceed the screening length.

The above arguments explain why a stable mosaic state can exists in *2D* liquids, but they are based on *2D*-specific factors (point-like dislocations with a self-screening logarithmic interaction). Due to these factors, the melting temperature of the crystal is too low to melt the local crystalline order in all clusters and only sufficient to create unbound dislocations and thus destroy the long-range order. In *3D* crystals, dislocations are lines or loops characterized by a large core energy and strong interactions. The melt of simple *3D* crystal (including the Lennard-Jones melt) is a simple liquid: the melting temperature in these crystals is sufficiently high to destroy the crystalline order on all length-scales. The temperature of a simple liquid can not be significantly lowered below the crystallization temperature. In contrast, glassforming liquids can be supercooled because nucleation and/or growth of new phase nuclei in these liquids can be dramatically slowed down. Then, the system may become a mosaic of slowly growing nuclei, but some additional factor is needed to stabilize the nuclei. In the *2D* liquid, this factor is the proliferation of dislocations limiting the size of a crystallite at temperatures above unbinding melting.

*3D* complex liquids are made of particles interacting via many-particle, non-central-force potentials. At sufficiently low temperatures, these interactions force particles to acquire low-energy relative positions at least in some small clusters. If these positions are compatible with crystalline order, the liquid crystallizes and no stable mosaic state appears. However, as the known examples of water and silicon dioxide molecules show, the bond lengths and bond angles corresponding to the minimum of the potential energy may have small differences from those in a crystal. We suggest that this weak incommensurability is a common feature of *3D* complex liquids. Then, particles in a small cluster can fluctuate near their minimum of energy positions while a crystalline order in a cluster of a larger size is possible only when the bonds and bond angles are stressed to fit into the lattice structure. The elastic energy of this stress diminishes the energy effect of crystallization, and thus the crystallization temperature, so the equilibrium or supercooled liquid can keep an almost crystalline local order in small clusters but incommensurability prevents these clusters from growing or coalescing. One can see here similarities with 2D liquids where instead of incommensurability the mosaic-stabilizing factor is the dislocation unbinding mechanism.

We believe that equilibrium mosaics exist in many liquids characterized by strong many-particle interactions. In some of these liquids, scattering experiments and other data show a detectible local order resembling that in crystals. The liquid-liquid phase transitions in pure liquids [27-31] are manifestations of local order that can change in a phase transition. Crystal-like local order has been found in liquid Benzene [32]. The small size of the ordered nuclei and the expected relatively small differences in internal energies between these nuclei and the surrounding less-ordered material can create challenging condition for experimental observation of the equilibrium mosaic state, especially a necessity to use unusually long equilibration times for the mosaic to form. To this end, we would like to draw attention to the behavior observed in liquid Benzene [33] and Quinolene [34] where scattering experiments using traditional equilibration times (hours) found at temperatures close to crystallization an increased data scatter, but some peaks in the temperature dependencies of small-scale characteristics when the equilibration times were an order of magnitude longer.

**Acknowledgments:** This work was supported by the Nonequilibrium Energy Research Center (NERC) which is an Energy Frontier Research Center funded by the U.S. Department of Energy, Office of Science, Office of Basic Energy Sciences, under Award Number DE-SC0000989.

**References**

1. I. Hamley, *Introduction to Soft Matter* (2nd edition), J. Wiley, Chichester, 2000.
2. Barrat J. L, Hansen J.P, *Basic concepts for simple and complex liquids*, Cambridge University Press, Cambridge, 2003.
3. R. G. Larson, *The Structure and Rheology of Complex Fluids (Topics in Chemical Engineering)*, Oxford University Press, New-York-Oxford, 1999.
4. W. M. Gelbart, A. Ben-Shaul, The "New" Science of "Complex Fluids", *J. Phys. Chem.* 1996, **100**, 13169–13189.
5. C. A. Angell, Glass-Formers and Viscous Liquid Slowdown since David Turnbull: Enduring Puzzles and New Twists, *MRS Bulletin* 2008, **33**, 544-555.
6. M. D. Ediger, Spatially Heterogeneous Dynamics in Supercooled Liquids, *Annu. Rev. Phys. Chem.* 2000, **51**, 99-128.
7. M. D. Ediger and J. L. Skinner, Single molecules rock and roll near the glass transition, *Science* 2001, **292**, 233-234.
8. L. D. Landau and E. M.Lifshitz, *Statistical Mechanics,* Reed, New York, 1980.
9. J.W Cahn, On spinodal decomposition, *Acta Met.* 1961, **9**, 795-804.
10. A. Z. Patashinskii and I. S.Jakub, Relaxation state in the neighborhood of a stratification point, *Sov. Solid State* 1976, **18**, 2114-2120.
11. R. Yamamoto and A. Onuki, Dynamics of highly supercooled liquids: Heterogeneity, rheology, and diffusion, *Phys. Rev. E* 1998, **58**, 3515-3529.
12. A. Z. Patashinski, M. A. Ratner, B. A. Grzybowski, R. Orlik, and A. C. Mitus, Heterogeneous Structure, Heterogeneous Dynamics, and Complex Behavior in Two-Dimensional Liquids, *J. Phys. Chem. Lett.* 2012, **3**, 2431-2435.
13. A. Z. Patashinski,R. Orlik, A. C. Mitus, B. A. Grzybowski, and M. A. Ratner, Melting in 2D: what type of phase transition? *J. Phys. Chem. C* 2011, **114**, 20749-20755.
14. D. R Nelson, B. I. Halperin, Dislocation-mediated melting in two dimensions, *Phys. Rev. B* 1979, **19**, 2457–2094.
15. K. J. Strandburg, Two-dimensional melting, *Rev. Mod. Phys.* 1988, **60**, 161–207.
16. H. Shiba, A. Onuki, T.Araki, Structural and dynamical heterogeneities in two- dimensional melting, *Europhys. Lett.* 2009, **86**, 66004-66007.
17. K. Chen, T. Kaplan, and M. Mostroller, Melting in Two-Dimensional Lennard-Jones Systems: Observation of a Metastable Hexatic Phase, *Phys. Rev. Lett.* 1995 **74**, 4019–4022.
18. K. Bagchi, H. C. Andersen, W. Swope, Observation of a two-stage melting transition in two dimensions, *Phys. Rev. E* 1996, **53**, 3794–3803.
19. P. J. Steinhard , D. R. Nelson, M. Ronchetti, Bond-orientational order in liquids and glasses, *Phys. Rev. B* 1983, **28**, 784-805.
20. P. J. Steinhard , D. R. Nelson, M. Ronchetti, Icosahedral Bond Orientational Order in Supercooled Liquids, *Phys. Rev. Lett.* 1983, **47**, 1297-1300.

21. A. C. Mitus, A. Z. Patashinski, A. Patrykiejew, and S. Sokolowski, Local structure, fluctuations, and freezing in two dimensions, *Phys. Rev. B* 2002, **66**, 1842021 -18420212.
22. A. C. Mitus, R. Orlik, and A. Z. Patashinski "Metrical vs. topological neighborhood relations and Lindemann melting criterion in two dimensions", Acta Phys. Polonica B 2004, **35**, 1501-1506.
23. J. Frenkel, *Kinetic Theory of Liquids* , ed. R. H. Fowler,P. Kapitza, N. F. Mott, Oxford University Press, Oxford 1947.
24. H. Eyring, Viscosity, Plasticity, and Diffusion as Examples of Absolute Reaction Rates, *J. Chem. Phys.* 1936, **4**, 283-292.
25. F. H.Stillinger, T. A. Weber, Inherent structure theory of liquids in the hard-sphere limit, *Science* 1984, **225**, 983-992.
26. A. Z. Patashinski, M. A. Ratner, Inherent amorphous structures and statistical mechanics of melting, *J. Chem. Phys.,* 1997, **106**, 7249-7256.
27. A. C. Mitus, A.Z. Patashinskii, B.I. Shumilo, "Liquid -liquid phase transition", *Phys. Lett. A* 1985, **113**, 41-44.
28. S. Harrington, R. Zhang, P. H. Poole, F. Sciortino, H. E. Stanley, Liquid-Liquid Phase Transition: Evidence from Simulations, *Phys. Rev. Lett.* 1997, **78**, 2409-2412.
29. O. Mishima and H. E. Stanley, "The Relationship between Liquid, Supercooled and Glassy Water," *Nature* 1998, **396**, 329-335.
30. L. Son, G. Rusakov, N. Katkov, Pressure-temperature phase diagrams of selenium and sulfur in terms of Patashinski model, *Physica A: Stat. Mech. and its Appl.* 2003, **324**, 634-644.
31. V. V. Brazhkin, Y. Katayama,. M. V. Kondrin, T. Hattori, A. G. Lyapin, and H. Saitoh, AsS melt under pressure: One substance, three liquids , *Phys. Rev. Lett.* 2008, **100**, 145701-145705.
32. A. H. Narten, Diffraction Pattern and Structure of Liquid Benzene, *J. Chem. Phys.* 1968, **48**, 1630-1635.
33. N. B. Rozhdestvenskaya, L.V. Smirnova, *J. Chem. Phys.* 1991, **195**, 1223-1230.
34. L. Letamendia, M. Belkari, O. Eloutassi, J. Rouch, D. Risso, P. Cordero, A. Z. Patashinski, Temperature anomalies of hypersound velocity and specific heat ratio in liquid Quinoline, *Physica A: Stat. Mech. and its Appl.* 2005, **354**, 34-40.

# Supplemental material

Mosaic states of the *2D* Lennard-Jones liquid are only found [13] in a narrow band of states in the *(T–ρ)* thermodynamic plane (see Fig. 1S).

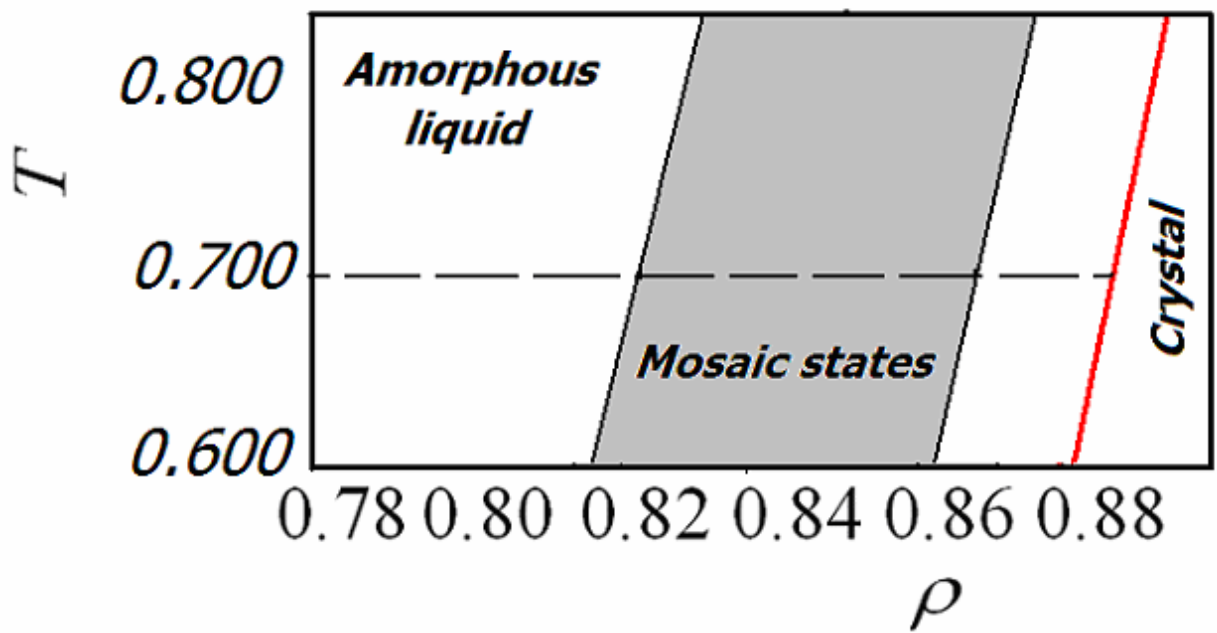

**Figure 6S. Mosaic states in the temperature-density thermodynamic plane.**

The boundaries of this band are approximately parallel to the melting (solidus) line that is close but outside of the band. Isotherms *T*=const and isochors *ρ*=const cross the band. Along an isotherm, the crystalline fraction of the liquid monotonously increases with increasing density $\rho$ as described in the main text (Fig. 2 and 3). For *T*=0.700, the *CL-AL* coarsened mosaics are shown in Fig. 2S. In the uniform liquid at *ρ*=0.60, small (3-5 particles) clusters of *CL* particles appear as rare and short-living fluctuations. At *ρ*=0.78, the liquid at any time includes ~10% of *CL* particles (Fig. 3), but the lifetimes of the small *CL* aggregates is of the order of particles vibration period in these aggregates [12,13]. Fig. 2S shows the coarsened *CL-AL* mosaic for T=0.700, $\rho$ =(0.82 – 0.87), in these pictures *CL* particles are represented by red and *AL* particles by yellow circles. At $\rho$<0.83, crystallites are spatially separated islands of *CL* particles in a matrix of *AA* particles (Fig. 2S, upper row); with $\rho$ increasing, both the sizes of largest crystallites and the fraction of *CL* particles monotonously increase. Between $\rho$=0.83 and $\rho$=0.84 the crystalline (and, correspondingly, the complimentary amorphous) fraction of the *2D* liquid reaches 50%, the percolation threshold for 2D random percolation. [1S].

In a finite system, the percolation point is replaced by a percolation range of densities [1S], so the percolation density $\rho_{perc}$ ~0.83-0.84 is rather a rough approximation. Typical large crystallites and also large amorphous clusters at densities close to-percolation include ~100 particles. At densities $\rho$>0.84, *CL* particles form a multi-connected crystalline matrix hosting islands of *AA* particles. At $\rho$=0.90 there are only few very small (less than 10 particles) *AL*-clusters representing vacancies and dislocations in the crystal.

A remarkable feature of the configurations in Fig. 2S is an approximate symmetry with regard to a simultaneous change *CL* ⇆ *AL* and $\delta\rho^* \leftrightarrows - \delta\rho^*$, $\delta\rho^*=(\rho^* -\rho^*_{perc}) <<\rho^*_{perc}$). The sampled configurations only qualitatively support the assumption of this *CL-AL* symmetry, to prove this assumption one needs to compare probabilities of corresponding configurations – a task not yet performed. The transformation here bears an apparent resemblance to the $m \leftrightarrows - m$, $h \leftrightarrows - h$ transformation for the classical ferromagnetic Ising model (with $m(r) =\pm 1$ as the local Ising variable and *h* the magnetic field). However, while in the Ising model the symmetry follows from the form of the Hamiltonian of the model, the *CL* ⇆ *AL* symmetry may be only approximate. This situation is generic in the Ising (scalar field φ(r)) universality class of critical behaviour: renormalization of the Hamiltonian to exclude the molecular-scale degrees-of-freedom (see, for example. [2S]) results in an effective Hamiltonian that parametrically depends on thermodynamic state.

In the mosaic range of densities $\rho$ = 0.82–0.85, an isolated cluster of *CL* or *AL* particles includes many tens of particles (see Fig.3). We assume that this size represents the elementary size *R* occupied by either structure. As discussed in the main text, the size of a *CL* cluster is limited by the screening length that is a function of density and temperature. An interesting but open question is long-range correlations between the positions of *CL* clusters near their percolation. To study these correlations and the percolation-related singularities (for example the expected singularity in the length of the *CL– AL* interface) one needs a system of a size much larger than the one studied here: systems of $N \sim 10^4$ particles are large enough to allow the measurements of the internal properties of clusters but too small to study long-range correlations between *CL* clusters.

## Supplemental References

1S. D. Stauffer, A. Aharony, *Introduction to Percolation Theory* (Taylor & Francis Ltd: London, 1994).

2S. A.Z. Patashinskii, V.L. Pokrovskii, *Fluctuation Theory of Phase Transitions* (Pergamon, Oxford 1979).